\documentclass[12pt,twoside]{article}
\usepackage{amsmath,epsfig}

\DeclareMathOperator*{\tsum}{{\textstyle \sum}}
\DeclareMathOperator*{\tprod}{{\textstyle \prod}}
\def\id{\mathrm{id}}

\makeatletter

\def\section{\@startsection{section}{1}{\z@}{-3.25ex plus -1ex minus
            -.2ex}{1.5ex plus .2ex}{\normalfont\bfseries}}

\def\thebibliography#1{\section*{\refname}\list
  {\@biblabel{\theenumiv}}{\settowidth\labelwidth{\@biblabel{#1}}%
    \leftmargin\labelwidth
    \advance\leftmargin\labelsep
    \footnotesize \parsep=0pt \itemsep=0pt
    \usecounter{enumiv}%
    \let\p@enumiv\@empty
    \def\theenumiv{\arabic{enumiv}}}%
    \def\newblock{\hskip .11em plus.33em minus.07em}%
    \sloppy\clubpenalty4000\widowpenalty4000
    \sfcode`\.=1000\relax}

\newlength\testa

\def\VL{\setlength{\testa}{1mm}\addtolength{\testa}{\arrayrulewidth}
             \hskip -2pt
             \vrule \@height \testa \@width 1mm \@depth -1mm 
             \vrule \@width  \arrayrulewidth \@height \testa \hskip 2pt}
\def\AL{\setlength{\testa}{1mm}\addtolength{\testa}{\arrayrulewidth}
             \hskip -2pt
             \vrule \@height \testa \@width 1mm \@depth -1mm
             \vrule \@width \arrayrulewidth \@depth -1mm \hskip 2pt}
\def\TL{\setlength{\testa}{1mm}\addtolength{\testa}{\arrayrulewidth}
             \hskip -2pt
             \vrule \@height \testa \@width 1mm \@depth -1mm
             \vrule \@width \arrayrulewidth \hskip 2pt}
\def\IL{\setlength{\testa}{1mm}\addtolength{\testa}{\arrayrulewidth}
             \hskip -2pt 
             \vrule \@height 0pt \@width 1mm \@depth 0pt 
             \vrule \@width \arrayrulewidth \hskip 2pt }

\def\ps@hepth{\addtolength{\headheight}{10pt}
                \addtolength{\topmargin}{-10pt}
      \def\@oddfoot{\hfil\thepage\hfil}\let\@evenfoot\@oddfoot
      \def\@evenhead{\hfil\footnotesize \begin{tabular}{r} 
       hep-th/9912220 \\ CPT-99/P.3784 	\end{tabular}}
       \let\@oddhead\@evenhead}

\makeatother

\allowdisplaybreaks[2]
\begin{document}

\thispagestyle{hepth}

\vskip 3mm
\begin{center}
{\renewcommand{\thefootnote}{\fnsymbol{footnote}}
{\large\bfseries On Feynman graphs as elements of a Hopf algebra}%
\footnote{pubished in: \emph{Quantum groups, noncommutative geometry
and fundamental physical interactions},
eds D. Kastler et al, Nova Science Publishers (1999) 233--242.}
\\[3ex]
{\scshape Raimar Wulkenhaar\footnote{e-mail: 
\texttt{raimar@cpt.univ-mrs.fr};
supported by the German Academic Exchange Service (DAAD), grant 
no.\ D/97/20386.}}
\setcounter{footnote}{0}}
\vskip 4mm

{\itshape Centre de Physique Th\'eorique \\
          CNRS - Luminy, Case 907 \\ 
          13288 Marseille Cedex 9, France} 

\end{center}
\vskip 4ex
 
\begin{abstract}
We review Kreimer's construction of a Hopf algebra associated to the
Feynman graphs of a perturbative quantum field theory.
\end{abstract}

\vskip 1ex

\section{Introduction}

This article is a companion to the contribution of Dirk Kreimer
\cite{k0} to this volume. With his discovery that divergent Feynman
graphs can be regarded as elements of a Hopf algebra whose antipode
implements renormalization \cite{k}, Kreimer has initiated a dynamic
development on the frontier of quantum field theory and noncommutative
geometry. In particular since Alain Connes and Dirk Kreimer \cite{ck}
established a structural link of this renormalization Hopf algebra to
a Hopf algebra found in the study of an index problem in
noncommutative geometry \cite{cm}, this topic belongs to the most
promising ones towards a unification of quantum field theory with
gravity. Connes and Kreimer realized that a subalgebra of the Hopf
algebra of renormalization (of a QFT given by a single divergent
Feynman graph) is isomorphic to the dual of the diffeomorphism group
of a manifold.  The central quest is now the search for the
counterpart of the entire renormalization Hopf algebra (of a realistic
QFT) replacing the diffeomorphism group. There is no doubt that the
latter object will deliver precious information on the short distance
structure of spacetime. Renormalization is our most powerful
microscope to see the smallest structures of the world!

Compared with these dreams, this article is extremely modest. We
review how the Hopf algebra is derived from Feynman graphs. This
review is intended to be pedagogical, we try to omit technical
details as far as possible and prefer to illustrate the essential
steps by typical examples from QED. The reader will find supplementary
information in the original papers \cite{k,kw,ck}. Our strategy is to
focus first on Feynman graphs without overlapping divergences (section
\ref{for}), because they yield a fairly simple Hopf algebra (section
\ref{hopf}). In section \ref{ov} we include overlapping divergences and
extend the Hopf algebra according to ideas developed in \cite{kw}. On
that level, the antipode of the Hopf algebra recovers the
combinatorics of renormalization. It reproduces the entire
renormalization if we allow for a deformation of the Hopf algebra by a
renormalization map $R$, see section \ref{R}. That map $R$ spoils
however the Hopf algebra axioms, and we hope to gain a deeper
understanding of this new structure in the future.

\section{Feynman graphs, parenthesized words and rooted trees}
\label{for}

Given a Feynman graph, let us draw boxes around all of its 
superficially (UV-) divergent sectors, for example 
\begin{equation}
\label{bsp}
\parbox{60mm}{\epsfig{file=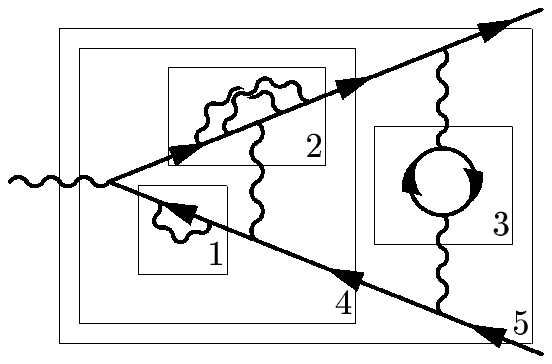}}\quad
= (((s_1)(v_2)v_4)(p_3)v_5)
\end{equation}
(As usual, straight lines stand for fermions and wavy lines for
bosons.) A criterion for superficial divergence of a region confined
in a box is power counting. If a box has $n_B$ bosonic and $n_F$
fermionic outgoing legs, the power counting degree of divergence $d$
is (in four dimensions) defined by $d:=4-n_B - \tfrac{3}{2} n_F \geq
0$.  Owing to symmetries the actual degree of divergence of one graph
or a sum of graphs can be lower than $d$, see ref.\ \cite{iz}.

Our example \eqref{bsp} is taken from a special class of Feynman
graphs which contain no overlapping divergences. This means that the
boxes can always be chosen non-intersecting. Any two boxes are,
therefore, either disjoint from each other, or one box is nested in
the other box. It is now convenient to rewrite such a Feynman graph in
a form where the relative position of the subdivergences is more
apparent. Starting with the set of innermost disjoint boxes we cut
them out of the graph but leave them in the next-larger box. In our
example, these are the boxes 1 and 2 which are cut out from the
graph in box 4. In box 4 we are therefore left with a graph with two
holes and two separate subgraphs flying around\footnote{When we cut
out a self-energy insertion, which means cutting a propagator into
two, we attach one of these new propagators to the self-energy graph
cut out. In this way we keep the number of holes in a graph
finite.}. Now we pass to the next larger box (no.~5) and cut out the
disjoint boxes from the graph. In the example we get a vertex graph
with two holes and the two boxes 3 and 4 flying around, where box 4
itself contains boxes 1 and 2. Now we replace the boxes by pairs of
opening-closing parentheses and order their contents such that the
mother graph stands on the right of its children graphs. This gives us
a ``parenthesized word'', PW for short.  The PW of our example
\eqref{bsp} looks as follows:
\begin{align}
&(((s_1)(v_2)v_4)(p_3)v_5) \\
&= \left(\left(\left(\parbox{17mm}{\epsfig{file=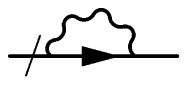}}\right)
\left(\parbox{20mm}{\epsfig{file=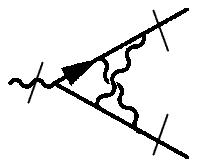}}\right)
\parbox{20mm}{\epsfig{file=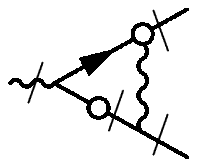}}\right)
\left(
\parbox{17mm}{\epsfig{file=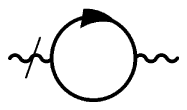}}\right)
\parbox{20mm}{\epsfig{file=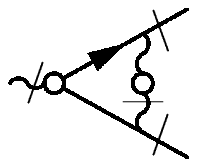}}\right) \notag
\end{align}
A slash through a propagator means amputation and a small circle
symbolizes a hole. We see that our building blocks are the Feynman
graphs with possible holes at any vertex and in any propagator.  In an
irreducible parenthesized word (iPW for short) the leftmost opening
parenthesis matches its rightmost closing parenthesis. A special type
of iPWs are the primitive PWs which contain no inner parentheses.

There is a second way of writing the same, which turns out to be the
adapted language in connection with noncommutative geometry. In an iPW
we call the rightmost graph (mother graph) the root. We connect to
this root the mother graphs of all its children boxes, and to these
mother graphs the mother graphs of their children boxes, and so on. In
this way we get a ``rooted tree'' whose vertices are labelled by
Feynman graphs with holes. The rooted tree of our example \eqref{bsp}
clearly looks as follows:
\begin{equation}
\parbox{85mm}{
\begin{picture}(80,40)
\put(45,30){\line(-1,-1){10}}
\put(34,19){$\bullet$}
\put(15,20){\epsfig{file=v4.eps}}
\put(45,30){\line(1,-1){10}}
\put(54,19){$\bullet$}
\put(58,15){\epsfig{file=p3.eps}}
\put(35,20){\line(-1,-1){10}}
\put(24,9){$\bullet$}
\put(10,-1){\epsfig{file=s1.eps}}
\put(35,20){\line(1,-1){10}}
\put(44,9){$\bullet$}
\put(48,0){\epsfig{file=v2.eps}}
\put(44,29){$\bullet$}
\put(44,29){\epsfig{file=v5.eps}}
\end{picture}
}=~ \parbox{25mm}{ 
\begin{picture}(20,33)
\put(14,19){$\bullet$}
\put(17,20){$v_5$}
\put(15,20){\line(-1,-1){5}}
\put(9,14){$\bullet$}
\put(5,16){$v_4$}
\put(15,20){\line(1,-1){5}}  
\put(19,14){$\bullet$}
\put(22,14){$p_3$}
\put(10,15){\line(-1,-1){5}}  
\put(4,9){$\bullet$}
\put(0,9){$s_1$}
\put(10,15){\line(1,-1){5}}
\put(14,9){$\bullet$}
\put(17,9){$v_2$}
\end{picture}}
\end{equation}
The tree is connected and simply connected for irreducible PWs and
disconnected for reducible PWs. The tree of a primitive PWs consists
solely of a labelled point.

\section{The Hopf algebra}
\label{hopf}

Dirk Kreimer has discovered \cite{k} that the PWs or rooted trees of
Feynman graphs form a Hopf algebra, whose antipode axiom reproduces
the forest formula of renormalization \cite{z}. Let us review these
ideas in some detail. We refer e.g.\ to appendix 1 in \cite{dk} for
a list of basic properties of Hopf algebras.

We consider the algebra $\mathcal{A}$ of polynomials over the rational
numbers $\mathbf{Q}$ of irreducible PWs resp.\ connected rooted
trees. 
%(So far we have excluded overlapping divergences, but
%they turn out to simply enlarge the algebra $\mathcal{A}$.) 
The multiplication in $\mathcal{A}$ is given by writing the trees
or PWs disjoint to each other. That multiplication is clearly
associative and commutative. We also adjoin a formal unit $e$ to
$\mathcal{A}$.

We are going to equip that algebra with the structure of a Hopf
algebra. The counit $\varepsilon: \mathcal{A} \to \mathbf{Q}$ is an
operation which annihilates trees resp.\ PWs:
\begin{align} 
\varepsilon[qe] &=q~,\quad q \in \mathbf{Q}, & \varepsilon[X] &=0~,\quad
\mathcal{A} \ni X \neq e~. 
\label{eps}
\end{align} 
The product being an assignment of one element to sums of pairs of other
elements, we expect the coproduct $\Delta: \mathcal{A} \to \mathcal{A}
\otimes \mathcal{A}$ to be the splitting of a given PW or tree into a sum of
pairs of PWs\,/\,trees. Before we give the details we have to define the
notion of a subword or subtree. A parenthesized subword (PSW for short) of a
PW is any of its iPW contained in it. In our example \eqref{bsp}, the
subwords of $(((s_1)(v_2)v_4)(p_3)v_5)$ are 
\begin{equation} 
(s_1)~,\quad
(v_2)~,\quad (p_3)~,\quad ((s_1)(v_2)v_4)~,\quad (((s_1)(v_2)v_4)(p_3)v_5)~.
\end{equation} 
The idea of the coproduct of a PW $X$ is that it returns as the left
factor of the tensor product any admissible product of PSWs
$X_i$ of $X$ and as the right factor that what remains when we remove
this left factor from $X$. A product $X_1 \cdots X_n$ is admissible if
any two $X_i,\,X_j$ contained in it do not intersect. For instance,
$(s_1)(p_3)$ or $(s_1)(v_2)$ are admissible, but
$(v_2)((s_1)(v_2)v_4)$ is not. We symbolize the removal of $X_1 \cdots
X_n$ from $X$ by $X/\tprod_{i=1}^n X_i$ (replaced by $0$ if
$X_1 \cdots X_n$ is not admissible). A special case is $X/X=e$. Now,
if the PSWs of the PW $X$ are $X_1,\dots,X_n$, we let $U$ be the set
of all (ordered) subsets of $\{1,\dots,n\}$ and define
\begin{align}
\Delta[e] &:= e\otimes e ~, \notag \\ 
\Delta[X] &:= e \otimes X + \tsum_U \Big\{ \tprod_{i \in U} \!\! X_i 
\otimes X/ \! \tprod_{i \in U} \!\! X_i \Big\}~. \label{delta_P}
\end{align}
For our example we find 
\begin{align}
\Delta\big[ (((s_1)(v_2)v_4)(p_3)v_5) \big] &=
e \otimes (((s_1)(v_2)v_4)(p_3)v_5) 
+ (s_1) \otimes ((v_2)v_4)(p_3)v_5) \notag 
\\
& + (v_2) \otimes ((s_1)v_4)(p_3)v_5)
+ (s_1)(v_2) \otimes ((v_4)(p_3)v_5) \notag 
\\
& + (p_3) \otimes (((s_1)(v_2)v_4)v_5) \notag 
+ (s_1)(p_3) \otimes (((v_2(v_4)v_5)
\\ 
& + (v_2)(p_3) \otimes (((s_1(v_4)v_5)
+ (s_1)(v_2)(p_3) \otimes ((v_4)v_5) \notag 
\\
& + ((s_1)(v_2)v_4)  \otimes ((p_3)v_5) 
+ ((s_1)(v_2)v_4) (p_3) \otimes (v_5) \notag 
\\
& + (((s_1)(v_2)v_4)(p_3)v_5) \otimes e ~.
\end{align}

In terms of rooted trees, the coproduct has an even more natural
interpretation. A subtree $T_i$ of a tree $T$ is what falls down if we
cut one edge of $T$, or it is the connected tree itself cut out of the
plane. The notion of an admissible product of PSWs finds its
counterpart in the set of admissible (multi-) cuts: the product $T_1
\cdots T_n$ is admissible iff on the path from any bottom vertex to
the root of $T$ we meet at most one of the $n$ cuts that have
produced the $T_i$. The rest $T/(T_1 \cdots T_n)$ is what remains
attached to the root if we cut away the subtrees $T_1,\dots,T_n$. The
analogue of \eqref{delta_P} is
\begin{equation}
\Delta[T] := e \otimes T + \tsum_C \Big\{ \tprod_{i \in U(C)} \!\!\!\! 
T_i  \otimes T/\!\!\! \tprod_{i \in U(C)} \!\!\! T_i \Big\}~, 
\label{delta_C}
\end{equation}
where each $C$ is an admissible multi-cut of $T$ which produces the
subtrees $\{T_i\}_{i \in U(C)}$. For our example \eqref{bsp} , the
coproduct reads in terms of trees
\begin{align}
\Delta \Bigg[ \;
\parbox{18mm}{
\begin{picture}(13,12)
\put(9,10){$\bullet$}
\put(12,11){$v_5$}
\put(10,11){\line(-1,-2){2.5}}
\put(6.5,5){$\bullet$}
\put(3,7){$v_4$}
\put(10,11){\line(1,-2){2.5}}
\put(11.5,5){$\bullet$}
\put(14,5){$p_3$}
\put(7.5,6){\line(-1,-2){2.5}}
\put(4,0){$\bullet$}
\put(0,0){$s_1$}
\put(7.5,6){\line(1,-2){2.5}}
\put(9,0){$\bullet$}
\put(11.5,0){$v_2$}
\end{picture}} 
\Bigg]
& = 
e \otimes 
\parbox{16mm}{
\begin{picture}(13,12)
\put(7,10){$\bullet$}
\put(10,11){$v_5$}
\put(8,11){\line(-1,-2){2.5}}
\put(4.5,5){$\bullet$}
\put(1,7){$v_4$}
\put(8,11){\line(1,-2){2.5}}
\put(9.5,5){$\bullet$}
\put(12,5){$p_3$}
\put(5.5,6){\line(-1,-2){2.5}}
\put(2,0){$\bullet$}
\put(-2,0){$s_1$}
\put(5.5,6){\line(1,-2){2.5}}
\put(7,0){$\bullet$}
\put(9.5,0){$v_2$}
\end{picture}} 
+ % s1
\parbox{4mm}{\begin{picture}(3,6)
\put(1,2){$\bullet$}
\put(1,5){$s_1$}
\end{picture}}
\otimes 
\parbox{16mm}{\begin{picture}(16,12)
\put(4,10){$\bullet$}
\put(7,11){$v_5$}
\put(5,11){\line(0,-1){5}}
\put(4,5){$\bullet$}
\put(0,7){$v_4$}
\put(5,11){\line(1,-1){5}}
\put(9,5){$\bullet$}
\put(12,5){$p_3$}
\put(5,6){\line(0,-1){5}}
\put(4,0){$\bullet$}
\put(0,0){$v_2$}
\end{picture}}
+ % v2
\parbox{4mm}{\begin{picture}(3,6)
\put(1,2){$\bullet$}
\put(1,5){$v_2$}
\end{picture}}
\otimes 
\parbox{16mm}{\begin{picture}(16,12)
\put(4,10){$\bullet$}
\put(7,11){$v_5$}
\put(5,11){\line(0,-1){5}}
\put(4,5){$\bullet$}
\put(0,7){$v_4$}
\put(5,11){\line(1,-1){5}}
\put(9,5){$\bullet$}
\put(12,5){$p_3$}
\put(5,6){\line(0,-1){5}}
\put(4,0){$\bullet$}
\put(0,0){$s_1$}
\end{picture}}
\notag \\ &+ % s1 v2
\parbox{9mm}{\begin{picture}(9,6)
\put(1,2){$\bullet$}
\put(1,5){$s_1$}
\put(6,2){$\bullet$}
\put(6,5){$v_2$}
\end{picture}}
\otimes 
\parbox{15mm}{\begin{picture}(15,8)
\put(6.5,5){$\bullet$}
\put(10,6){$v_5$}
\put(7.5,6){\line(-1,-2){2.5}}
\put(4,0){$\bullet$}
\put(0,0){$v_4$}
\put(7.5,6){\line(1,-2){2.5}}
\put(9,0){$\bullet$}
\put(12,0){$p_3$}
\end{picture}}
+ % p3
\parbox{4mm}{\begin{picture}(3,6)
\put(1,2){$\bullet$}
\put(1,5){$p_3$}
\end{picture}}
\otimes 
\parbox{11mm}{\begin{picture}(8,12)
\put(3,10){$\bullet$}
\put(6,11){$v_5$}
\put(4,11){\line(0,-1){5}}
\put(3,5){$\bullet$}
\put(6,6){$v_4$}
\put(4,6){\line(-1,-2){2.5}}
\put(0.5,0){$\bullet$}
\put(-3,0){$s_1$}
\put(4,6){\line(1,-2){2.5}}
\put(5.5,0){$\bullet$}
\put(8,0){$v_2$}
\end{picture}}
+ % p3 s1
\parbox{10mm}{\begin{picture}(9,6)
\put(1,2){$\bullet$}
\put(1,5){$s_1$}
\put(6,2){$\bullet$}
\put(6,5){$p_3$}
\end{picture}}
\otimes 
\parbox{8mm}{\begin{picture}(5,12)
\put(0,10){$\bullet$}
\put(3,10){$v_5$}
\put(1,11){\line(0,-1){5}}
\put(0,5){$\bullet$}
\put(3,5){$v_4$}
\put(1,6){\line(0,-1){5}}
\put(0,0){$\bullet$}
\put(3,0){$v_2$}
\end{picture}}
\notag \\ & + % p3 v2
\parbox{10mm}{\begin{picture}(9,6)
\put(1,2){$\bullet$}
\put(1,5){$v_2$}
\put(6,2){$\bullet$}
\put(6,5){$p_3$}
\end{picture}}
\otimes 
\parbox{8mm}{\begin{picture}(5,12)
\put(0,10){$\bullet$}
\put(3,10){$v_5$}
\put(1,11){\line(0,-1){5}}
\put(0,5){$\bullet$}
\put(3,5){$v_4$}
\put(1,6){\line(0,-1){5}}
\put(0,0){$\bullet$}
\put(3,0){$s_1$}
\end{picture}}
+ % p3 s1 v2
\parbox{15mm}{\begin{picture}(12,6)
\put(1,2){$\bullet$}
\put(1,5){$s_1$}
\put(6,2){$\bullet$}
\put(6,5){$v_2$}
\put(11,2){$\bullet$}
\put(11,5){$p_3$}
\end{picture}}
\otimes 
\parbox{8mm}{\begin{picture}(5,7)
\put(0,5){$\bullet$}
\put(3,5){$v_5$}
\put(1,6){\line(0,-1){5}}
\put(0,0){$\bullet$}
\put(3,0){$v_4$}
\end{picture}}
+ % s1 v2 v4 
\parbox{12mm}{\begin{picture}(10,7)
\put(5,5){$\bullet$}
\put(8,6){$v_4$}
\put(6,6){\line(-1,-2){2.5}}
\put(2.5,0){$\bullet$}
\put(-1,0){$s_1$}
\put(6,6){\line(1,-2){2.5}}
\put(7.5,0){$\bullet$}
\put(10,0){$v_2$}
\end{picture}}
\otimes
\parbox{8mm}{\begin{picture}(5,7)
\put(0,5){$\bullet$}
\put(3,5){$v_5$}
\put(1,6){\line(0,-1){5}}
\put(0,0){$\bullet$}
\put(3,0){$p_3$}
\end{picture}}
\notag \\ & + % s1 v2 v4 p3
\parbox{13mm}{\begin{picture}(10,7)
\put(5,5){$\bullet$}
\put(8,6){$v_4$}
\put(6,6){\line(-1,-2){2.5}}
\put(2.5,0){$\bullet$}
\put(-1,0){$s_1$}
\put(6,6){\line(1,-2){2.5}}
\put(7.5,0){$\bullet$}
\put(10,0){$v_2$}
\end{picture}}
\parbox{4mm}{\begin{picture}(3,6)
\put(1,2){$\bullet$}
\put(1,5){$p_3$}
\end{picture}}
\otimes
\parbox{4mm}{\begin{picture}(3,6)
\put(1,2){$\bullet$}
\put(1,5){$v_5$}
\end{picture}}
+
\parbox{16mm}{
\begin{picture}(13,12)
\put(7,10){$\bullet$}
\put(10,11){$v_5$}
\put(8,11){\line(-1,-2){2.5}}
\put(4.5,5){$\bullet$}
\put(1,7){$v_4$}
\put(8,11){\line(1,-2){2.5}}
\put(9.5,5){$\bullet$}
\put(12,5){$p_3$}
\put(5.5,6){\line(-1,-2){2.5}}
\put(2,0){$\bullet$}
\put(-2,0){$s_1$}
\put(5.5,6){\line(1,-2){2.5}}
\put(7,0){$\bullet$}
\put(9.5,0){$v_2$}
\end{picture}} 
\otimes e ~. \label{delta_T}
\end{align}
The coproduct is coassociative,
\begin{equation}
(\Delta \otimes \id) \circ \Delta[X] = X 
= (\id \otimes \Delta) \circ \Delta[X]~. \label{coass}
\end{equation}
If we split a PW into a sum of two, it is the same to split the right
or the left factors further. The proof of \eqref{coass} is
non-trivial, it can be performed by induction \cite{k,ck} or directly 
\cite{kw}. The coproduct is, however, not cocommutative, which means
that in general 
\[
\tau \circ \Delta[X] \neq \Delta[X]~.
\]
Here, $\tau [X \otimes Y] = Y \otimes X$ is the flip operator. 

From \eqref{eps} and \eqref{delta_P} the following relation between
counit and coproduct is obvious:
\begin{equation}
(\varepsilon \otimes \id) \circ \Delta[X] = 
(\id \otimes \varepsilon) \circ \Delta[X] = X~. \label{coalg}
\end{equation}
Indeed, if we consider a monomial $X= \tprod_i X_i$ of iPWs resp.\
connected trees, we get $\Delta[X] = e \otimes X + \tsum Z \otimes Z'
+ X \otimes e$, where $\tsum Z \otimes Z'$ represents all terms which
do not contain the unit $e$ and which are therefore annihilated by
$\varepsilon$. The identity \eqref{coalg} means that our algebra
$\mathcal{A}$ is also a coalgebra. It is even a bialgebra as the
algebra and coalgebra operations are compatible. Denoting the
multiplication in $\mathcal{A}$ by $m$ we have 
\begin{equation}
\Delta \circ m[X \otimes Y] = (m \otimes m) \circ 
(\id \otimes \tau \otimes \id) \circ (\Delta \otimes \Delta) 
[X \otimes Y]~,
\end{equation}
due to the fact that the subwords of a product word $XY$ are the
subwords of $X$ and the subwords of $Y$ together. 

The bialgebra $\mathcal{A}$ also has an antipode $S: \mathcal{A} \to
\mathcal{A}$ which makes it to a Hopf algebra. It is on PWs and rooted
trees recursively defined by
\begin{align}
S[e]  &= e ~,\notag \\
S[XY] &= S[Y] S[X] ~,\qquad \forall X,Y \in \mathcal{A}~, \notag
\\
S[X]  &= -X - m \circ (S \otimes \id) \circ P_2 \Delta [X]~,
\qquad \forall \;\mathrm{iPW}~X \in \mathcal{A}~, \label{anti}
\end{align}
where $P_2: \mathcal{A} \otimes \mathcal{A} \to \mathcal{A} \otimes
\mathcal{A}$ is an operation annihilating all terms which contain the
unit $e$. Note that $S[X]=-X$ if $X$ is primitive. On rooted trees we
can give a more natural definition of the antipode. In $P_2 \Delta$ we
take precisely the proper admissible cuts into account, proper in the
sense that the cut of the entire tree out of the plane is not
included. Since the recursive definition of $S$ is translated into a
recursive application of $P_2 \Delta$ to the leftmost factor, it is
not difficult to see \cite{ck} that \eqref{anti} is equivalent to
\begin{equation}
S[X] = -X - \tsum_{C_a} \Big\{ (-1)^{\#(C_a)} 
\big\{ \hskip -0.5ex \tprod_{i \in U(C_a)} \hskip -0.8em X_i \big\} \, 
\big\{ X/\hskip -0.8em \tprod_{i \in U(C_a)} \hskip -0.8em X_i \big\} 
\Big\} ~.
\label{stree}
\end{equation}
Now, the sum runs over all proper multi-cuts $C_a$ consisting of
$\#(C_a)$ single cuts, where each $C_a$ cuts away the trees $X_i$, $i
\in U(C_a)$ which need no longer to be subtrees of $X$ in the previous
sense. It remains the tree $X/\tprod_{i \in U(C_a)} X_i$ which
contains the root. The last line of \eqref{anti} can equivalently be
written as
\begin{equation}
S[X]  = -X - m \circ (\id \otimes S) \circ P_2 \Delta [X]~,
\tag{$\ref{anti}'$} \label{anti'}
\end{equation}
The easiest way to see this is to realize that both versions yield the
same formula \eqref{stree}.

It is probably a good idea to illustrate the antipode for our example,
although it is a bit lengthy:
\begin{align}
S \Bigg[ \;
\parbox{18mm}{
\begin{picture}(13,12)
\put(9,10){$\bullet$}
\put(12,11){$v_5$}
\put(10,11){\line(-1,-2){2.5}}
\put(6.5,5){$\bullet$}
\put(3,7){$v_4$}
\put(10,11){\line(1,-2){2.5}}
\put(11.5,5){$\bullet$}
\put(14,5){$p_3$}
\put(7.5,6){\line(-1,-2){2.5}}
\put(4,0){$\bullet$}
\put(0,0){$s_1$}
\put(7.5,6){\line(1,-2){2.5}}
\put(9,0){$\bullet$}
\put(11.5,0){$v_2$}
\end{picture}} 
\Bigg]
& = 
- \parbox{16mm}{
\begin{picture}(13,12)
\put(7,10){$\bullet$}
\put(10,11){$v_5$}
\put(8,11){\line(-1,-2){2.5}}
\put(4.5,5){$\bullet$}
\put(1,7){$v_4$}
\put(8,11){\line(1,-2){2.5}}
\put(9.5,5){$\bullet$}
\put(12,5){$p_3$}
\put(5.5,6){\line(-1,-2){2.5}}
\put(2,0){$\bullet$}
\put(-2,0){$s_1$}
\put(5.5,6){\line(1,-2){2.5}}
\put(7,0){$\bullet$}
\put(9.5,0){$v_2$}
\end{picture}} 
+ % s1
\parbox{6mm}{\begin{picture}(3,6)
\put(1,2){$\bullet$}
\put(1,5){$s_1$}
\end{picture}}
\parbox{16mm}{\begin{picture}(16,12)
\put(4,10){$\bullet$}
\put(7,11){$v_5$}
\put(5,11){\line(0,-1){5}}
\put(4,5){$\bullet$}
\put(0,7){$v_4$}
\put(5,11){\line(1,-1){5}}
\put(9,5){$\bullet$}
\put(12,5){$p_3$}
\put(5,6){\line(0,-1){5}}
\put(4,0){$\bullet$}
\put(0,0){$v_2$}
\end{picture}}
+ % v2
\parbox{6mm}{\begin{picture}(3,6)
\put(1,2){$\bullet$}
\put(1,5){$v_2$}
\end{picture}}
\parbox{16mm}{\begin{picture}(16,12)
\put(4,10){$\bullet$}
\put(7,11){$v_5$}
\put(5,11){\line(0,-1){5}}
\put(4,5){$\bullet$}
\put(0,7){$v_4$}
\put(5,11){\line(1,-1){5}}
\put(9,5){$\bullet$}
\put(12,5){$p_3$}
\put(5,6){\line(0,-1){5}}
\put(4,0){$\bullet$}
\put(0,0){$s_1$}
\end{picture}}
- % s1 v2
\parbox{9mm}{\begin{picture}(9,6)
\put(1,2){$\bullet$}
\put(1,5){$s_1$}
\put(6,2){$\bullet$}
\put(6,5){$v_2$}
\end{picture}}
\parbox{15mm}{\begin{picture}(15,8)
\put(6.5,5){$\bullet$}
\put(9.5,6){$v_5$}
\put(7.5,6){\line(-1,-2){2.5}}
\put(4,0){$\bullet$}
\put(0,0){$v_4$}
\put(7.5,6){\line(1,-2){2.5}}
\put(9,0){$\bullet$}
\put(12,0){$p_3$}
\end{picture}}
\notag \\[1ex] & + % p3
\parbox{4mm}{\begin{picture}(3,6)
\put(1,2){$\bullet$}
\put(1,5){$p_3$}
\end{picture}}
\parbox{11mm}{\begin{picture}(8,12)
\put(3,10){$\bullet$}
\put(6,11){$v_5$}
\put(4,11){\line(0,-1){5}}
\put(3,5){$\bullet$}
\put(6,6){$v_4$}
\put(4,6){\line(-1,-2){2.5}}
\put(0.5,0){$\bullet$}
\put(-3,0){$s_1$}
\put(4,6){\line(1,-2){2.5}}
\put(5.5,0){$\bullet$}
\put(8,0){$v_2$}
\end{picture}}
- % p3 s1
\parbox{11mm}{\begin{picture}(9,6)
\put(1,2){$\bullet$}
\put(1,5){$s_1$}
\put(6,2){$\bullet$}
\put(6,5){$p_3$}
\end{picture}}
\parbox{8mm}{\begin{picture}(5,12)
\put(0,10){$\bullet$}
\put(3,10){$v_5$}
\put(1,11){\line(0,-1){5}}
\put(0,5){$\bullet$}
\put(3,5){$v_4$}
\put(1,6){\line(0,-1){5}}
\put(0,0){$\bullet$}
\put(3,0){$v_2$}
\end{picture}}
- % p3 v2
\parbox{11mm}{\begin{picture}(9,6)
\put(1,2){$\bullet$}
\put(1,5){$v_2$}
\put(6,2){$\bullet$}
\put(6,5){$p_3$}
\end{picture}}
\parbox{8mm}{\begin{picture}(5,12)
\put(0,10){$\bullet$}
\put(3,10){$v_5$}
\put(1,11){\line(0,-1){5}}
\put(0,5){$\bullet$}
\put(3,5){$v_4$}
\put(1,6){\line(0,-1){5}}
\put(0,0){$\bullet$}
\put(3,0){$s_1$}
\end{picture}}
+ % p3 s1 v2
\parbox{16mm}{\begin{picture}(12,6)
\put(1,2){$\bullet$}
\put(1,5){$s_1$}
\put(6,2){$\bullet$}
\put(6,5){$v_2$}
\put(11,2){$\bullet$}
\put(11,5){$p_3$}
\end{picture}}
\parbox{8mm}{\begin{picture}(5,7)
\put(0,5){$\bullet$}
\put(3,5){$v_5$}
\put(1,6){\line(0,-1){5}}
\put(0,0){$\bullet$}
\put(3,0){$v_4$}
\end{picture}}
\\[2ex] & + % s1 v2 v4 
\parbox{14mm}{\begin{picture}(10,7)
\put(5,5){$\bullet$}
\put(8,6){$v_4$}
\put(6,6){\line(-1,-2){2.5}}
\put(2.5,0){$\bullet$}
\put(-1,0){$s_1$}
\put(6,6){\line(1,-2){2.5}}
\put(7.5,0){$\bullet$}
\put(10,0){$v_2$}
\end{picture}}
\parbox{8mm}{\begin{picture}(5,7)
\put(0,5){$\bullet$}
\put(3,5){$v_5$}
\put(1,6){\line(0,-1){5}}
\put(0,0){$\bullet$}
\put(3,0){$p_3$}
\end{picture}}
- % s1 - v2+v4 - v5+p3
\parbox{6mm}{\begin{picture}(3,6)
\put(1,2){$\bullet$}
\put(1,5){$s_1$}
\end{picture}}
\parbox{8mm}{\begin{picture}(5,7)
\put(0,5){$\bullet$}
\put(3,5){$v_4$}
\put(1,6){\line(0,-1){5}}
\put(0,0){$\bullet$}
\put(3,0){$v_2$}
\end{picture}}
\parbox{8mm}{\begin{picture}(5,7)
\put(0,5){$\bullet$}
\put(3,5){$v_5$}
\put(1,6){\line(0,-1){5}}
\put(0,0){$\bullet$}
\put(3,0){$p_3$}
\end{picture}}
- % v2 - s1+v4 - v5+p3
\parbox{6mm}{\begin{picture}(3,6)
\put(1,2){$\bullet$}
\put(1,5){$v_2$}
\end{picture}}
\parbox{8mm}{\begin{picture}(5,7)
\put(0,5){$\bullet$}
\put(3,5){$v_4$}
\put(1,6){\line(0,-1){5}}
\put(0,0){$\bullet$}
\put(3,0){$s_1$}
\end{picture}}
\parbox{8mm}{\begin{picture}(5,7)
\put(0,5){$\bullet$}
\put(3,5){$v_5$}
\put(1,6){\line(0,-1){5}}
\put(0,0){$\bullet$}
\put(3,0){$p_3$}
\end{picture}}
+ % s1 - v2 - v4 - v5+p3
\parbox{16mm}{\begin{picture}(12,6)
\put(1,2){$\bullet$}
\put(1,5){$s_1$}
\put(6,2){$\bullet$}
\put(6,5){$v_2$}
\put(11,2){$\bullet$}
\put(11,5){$v_4$}
\end{picture}}
\parbox{8mm}{\begin{picture}(5,7)
\put(0,5){$\bullet$}
\put(3,5){$v_5$}
\put(1,6){\line(0,-1){5}}
\put(0,0){$\bullet$}
\put(3,0){$p_3$}
\end{picture}}
\notag \\[2ex] & - % s1+v2+v4 - p3 - p5
\parbox{13mm}{\begin{picture}(10,7)
\put(4.5,5){$\bullet$}
\put(7.5,6){$v_4$}
\put(5.5,6){\line(-1,-2){2.5}}
\put(2,0){$\bullet$}
\put(-1.5,0){$s_1$}
\put(5.5,6){\line(1,-2){2.5}}
\put(7,0){$\bullet$}
\put(9.5,0){$v_2$}
\end{picture}}
\parbox{9mm}{\begin{picture}(8,6)
\put(1,2){$\bullet$}
\put(1,5){$p_3$}
\put(6,2){$\bullet$}
\put(6,5){$v_5$}
\end{picture}}
+ % s1 - v2+v4 - v5 - p3
\parbox{6mm}{\begin{picture}(3,6)
\put(1,2){$\bullet$}
\put(1,5){$s_1$}
\end{picture}}
\parbox{8mm}{\begin{picture}(5,7)
\put(0,5){$\bullet$}
\put(3,5){$v_4$}
\put(1,6){\line(0,-1){5}}
\put(0,0){$\bullet$}
\put(3,0){$v_2$}
\end{picture}}
\parbox{9mm}{\begin{picture}(8,6)
\put(1,2){$\bullet$}
\put(1,5){$p_3$}
\put(6,2){$\bullet$}
\put(6,5){$v_5$}
\end{picture}}
+ % v2 - s1+v4 - v5 - p3
\parbox{6mm}{\begin{picture}(3,6)
\put(1,2){$\bullet$}
\put(1,5){$s_1$}
\end{picture}}
\parbox{8mm}{\begin{picture}(5,7)
\put(0,5){$\bullet$}
\put(3,5){$v_4$}
\put(1,6){\line(0,-1){5}}
\put(0,0){$\bullet$}
\put(3,0){$v_2$}
\end{picture}}
\parbox{9mm}{\begin{picture}(8,6)
\put(1,2){$\bullet$}
\put(1,5){$p_3$}
\put(6,2){$\bullet$}
\put(6,5){$v_5$}
\end{picture}}
-
\parbox{9mm}{\begin{picture}(8,6)
\put(1,2){$\bullet$}
\put(1,5){$s_1$}
\put(6,2){$\bullet$}
\put(6,5){$v_2$}
\put(11,2){$\bullet$}
\put(11,5){$p_3$}
\put(16,2){$\bullet$}
\put(16,5){$v_4$}
\put(21,2){$\bullet$}
\put(21,5){$v_5$}
\end{picture}}
\notag
\end{align}

We can now check the antipode axioms
\begin{align}
m \circ (S \otimes \id) \circ \Delta [e] &=  
m \circ (\id \otimes S) \circ \Delta [e] = e ~,\notag 
\\
m \circ (S \otimes \id) \circ \Delta [X] &=  
m \circ (\id \otimes S) \circ \Delta [X] = 0 \qquad \forall X \neq e ~.
\label{hop}
\end{align}
The second line is an immediate consequence of \eqref{anti},
\eqref{anti'} and of the identity
$\Delta[X] = e \otimes X + X \otimes e + P_2 \Delta[X]$.

The reader may worry what all that has to do with Feynman graphs. What
we have only used are parenthesized words or rooted trees whose
building blocks are letters of some alphabet. Indeed, it was pointed
out in \cite{k3} that the Hopf algebra structure is based on
elementary set theoretical considerations. Feynman graphs are just an
example. There is however an important observation, due to Dirk
Kreimer \cite{k}, which makes the application of these set theoretic
tools tremendously important for Feynman graphs. For $X$ being the
parenthesized word or rooted tree of a Feynman graph, the antipode $S$
of \eqref{anti} reproduces precisely the combinatorics of the forest
formula which governs the renormalization of perturbative QFTs. We will
return to this achievement in section~\ref{R}.

\section{Overlapping divergences}
\label{ov}

Feynman graphs may contain overlapping divergences which at first
sight do not fit into the language of parenthesized words or rooted
trees. However, it turns out that an overlapping divergence can be
represented by a linear combination of words or trees, where
additional primitive elements arise. We present here a particular
construction \cite{kw} of this linear combination. For this purpose we
enlarge the class of PWs by including words with several lines -- one
line for each maximal forest. We extend the Hopf algebra operations to
this larger class and show that they give rise to a linear combination
of one-line PWs and new primitive elements.

Let us consider the following graph borrowed from QED:
\[
\parbox{24mm}{\epsfig{file=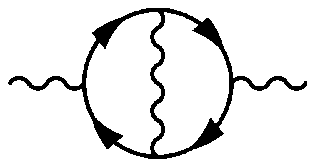}}
\]
There is no possibility to draw non-intersecting boxes around all
superficially divergent sectors. We can however try to draw
non-intersecting boxes around a maximal number of superficial
divergences, which will be possible in several ways. In our example we
have two possibilities:
\begin{equation}
\parbox{31mm}{
\begin{picture}(31,24)
\put(0,11.5){\parbox{30mm}{\epsfig{file=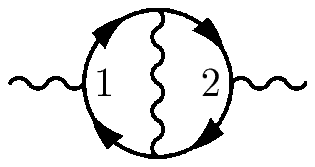}}}
\put(6,3){\rule{0.2pt}{19mm}} %ver
\put(6,3){\rule{11mm}{0.2pt}} %hor
\put(17,3){\rule{0.2pt}{19mm}} %ver
\put(6,22){\rule{11mm}{0.2pt}} %hor
  \put(4,1){\rule{0.2pt}{23mm}} %ver
  \put(4,1){\rule{21mm}{0.2pt}} %hor
  \put(25,1){\rule{0.2pt}{23mm}} %ver
  \put(4,24){\rule{21mm}{0.2pt}} %hor
\end{picture}} = ((v_1)p_2)
\qquad
\mbox{or}
\qquad
\parbox{31mm}{
\begin{picture}(31,24)
\put(0,11.5){\parbox{30mm}{\epsfig{file=polar.eps}}}
\put(13,3){\rule{0.2pt}{19mm}} %ver
\put(13,3){\rule{11mm}{0.2pt}} %hor
\put(24,3){\rule{0.2pt}{19mm}} %ver
\put(13,22){\rule{11mm}{0.2pt}} %hor
  \put(5,1){\rule{0.2pt}{23mm}} %ver
  \put(5,1){\rule{21mm}{0.2pt}} %hor
  \put(26,1){\rule{0.2pt}{23mm}} %ver
  \put(5,24){\rule{21mm}{0.2pt}} %hor
\end{picture}}
= ((v_2)p_1)~.
\label{PV0}
\end{equation}
We have resolved the overlapping divergence into (here two) maximal
forests. A forest of a Feynman graph is a set of
one-particle-irreducible (the graph remains connected after cutting an
arbitrary line) divergent subgraphs which do not overlap. A maximal
forest for a given Feynman graph is a forest which is not contained in
any other forest of that graph.

Our idea is now to bundle the parenthesized words of the $n$ maximal
forests of a given graph to an $n$-line PW. We write the single PWs of
each maximal forest as different rows and connect by a tree of lines
the closing parentheses of identical boxes occurring in different
rows. The PW of our example is 
\begin{equation}
\begin{array}{rl} 
((v_1) p_2) &\VL \\ 
((v_2) p_1) &\AL
\end{array} ~,
\label{PV}
\end{equation}
because the outermost parentheses of both rows represent the same
large box in \eqref{PV0}. Other examples are 
\begin{align}
& \parbox{40mm}{\epsfig{file=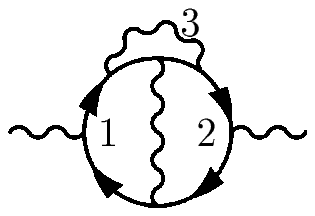}} && 
\begin{array}{rrrlrrl}
(((v_3) &\VL& v_{13}) p_2) &\VL \\
(((v_3) &\AL& v_{23}) p_1) &\AL
\end{array} 
\label{PVV}
\\[1ex]
& \parbox{40mm}{\epsfig{file=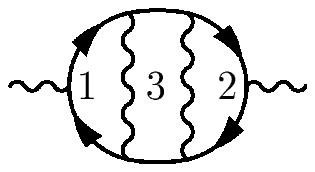}} && 
\begin{array}{lrrrr}
( &(v_1) \VL&(v_2) \VL&         & p_3) \VL \\
((&(v_1) \AL&      \IL& v_{13}) & p_2) \TL \\
((&         &(v_2) \AL& v_{23}) & p_1) \AL
\end{array}
\label{VVP}
\end{align}
All we have to change now is the notion of the parenthesized subword
(PSW) and of its removal. A PSW $X_i$ of $X$ is everything between a
set of connected closing parentheses and its matching opening
parentheses. Disconnected rows of $X$ which are accidentally between
connected rows are not part of the PSW $X$ under consideration, and
identical rows are condensed to one copy. Thus, apart from the total
PW, the proper PSWs in the examples are

\begin{align}
&\eqref{PV}:  &&  (v_1)\,,~(v_2) \notag \\
&\eqref{PVV}: &&  (v_3)\,,~((v_3)v_{13})\,,~((v_3)v_{23})\,,~ 
\label{suwo} \\
&\eqref{VVP}: &&  (v_1)\,,~(v_2)\,,~ ((v_1)v_{13})\,,~((v_2)v_{23})\,.
\notag
\end{align}
The removal of a product $\prod_i X_i$ of PSWs of $X$ from $X$ is
defined as follows: If $\prod_i X_i =X$ we define
$X/X=e$. Otherwise we label the rows of $X$. We give to the $X_i$-rows the
labels of the $X$-rows they are contained in. We delete
from $X$ and all $X_i$ all but those rows whose labels occur in each
of the chosen PSWs $X_i$. Let the results be $X'$ and $X_i'$. If there 
remains no row at all or if $X_i' \cap X_j' \neq \emptyset$ for some
pair $\{X_i',X_j'\}$ then we put $X/\prod_i X_i=0$.
Otherwise $X/\prod X_i$ is given by removing all $X_i'$ from $X'$.

With these modifications, the formulae \eqref{eps}, \eqref{delta_P}
and \eqref{anti} define counit, coproduct and antipode of a Hopf
algebra, and the properties \eqref{coass}, \eqref{coalg} and
\eqref{hop} remain unchanged, see \cite{kw}. The antipode reproduces
now the combinatorics of the forest formula for any Feynman graph. 

There is a way to return to the one-line PWs or rooted trees. It
suffices to define a ``primitivator'' $\mathcal{P}$ which maps
overlapping divergences to primitive elements. Let $X$ be an iPW with
proper PSWs $X_i \neq X$, $i=1,\dots,n$, and $U \subset
\{1,\dots,n\}$. Let us write the exterior parentheses of iPWs
explicitly, i.e.\ $(X)$ instead of $X$ and $(X_i)$ instead of $X_i$
and $(\mathcal{P}[X/\tprod_{i \in U} X_i])$ instead of
$\mathcal{P}[X/\tprod_{i \in U} X_i]$. With this convention we define
\begin{equation}
\mathcal{P}[(X)] := (X) - \tsum_{U} \Big( \tprod_{i \in U} (X_i) \:   
\mathcal{P}\big[ X/\tprod_{i \in U} X_i\big] \Big)~.
\label{Prim}
\end{equation}
We have proved in \cite{kw} that $\mathcal{P}[(X)]$ is primitive in the
following sense:
\begin{equation}   
\Delta[\mathcal{P}[(X)]]= e \otimes \mathcal{P}[(X)] 
+ \mathcal{P}[(X)] \otimes e ~.
\label{DP}
\end{equation}
If $(X)$ is primitive it contains no PSWs. Hence we have $U =
\emptyset$ and $\mathcal{P}[(X)]=(X)$. For $(X)$ and $(Y)$ being
primitive we compute $\mathcal{P}[((Y)X)] = ((Y)X)-((Y)X)=0$. By
induction it is easy to show that $\mathcal{P}[Y]=0$ for any
non-primitive one-line iPW $Y$. On the other hand, $\mathcal{P}[(X)]
\neq 0$ if $(X)$ is an overlapping divergence, and we can replace $X$
by the linear combination $\mathcal{P}[(X)] + \tsum_{U} \Big(
\tprod_{i \in U} (X_i) \: \mathcal{P}\big[ X/\tprod_{i \in U} X_i\big]
\Big)$.  If $(X)$ contains no overlapping subdivergences, all $X_i$
are one-line PWs. Since the $\mathcal{P}[(X)]$ form additional
primitive (i.e.\ one-line) elements of the Hopf algebra, we have
written the multi-line overlapping divergence $(X)$ as a linear
combination of one-line PWs. In other words, our Hopf algebra of
arbitrary Feynman graphs is isomorphic to a Hopf algebra of one-line
PWs, and this is precisely Kreimer's original Hopf algebra
\cite{k}. The primitive elements of Kreimer's Hopf algebra are the
graphically primitive elements and from each overlapping divergence a
computational-primitive element. We have given an explicit
construction of the latter. The same can be achieved, for instance, by
Schwinger-Dyson techniques \cite{k} or set theoretic considerations
\cite{k3}. 

Let us evaluate the primitivators of our examples, thereby giving the
decomposition into rooted trees. Using \eqref{suwo} we get 
\begin{align*}
o_{1,2|2,1} := \mathcal{P}\bigg[ 
\begin{array}{rl} 
((v_1) p_2) &\VL \\ 
((v_2) p_1) &\AL
\end{array} \bigg] & = \begin{array}{rl} 
((v_1) p_2) &\VL \\ 
((v_2) p_1) &\AL
\end{array} - ((v_1) p_2) - ((v_2) p_1) ~,
\\
\mathcal{P} \bigg[\begin{array}{rlrl} 
(((v_3) &\VL& v_{13}) p_2) &\VL \\
(((v_3) &\AL& v_{23}) p_1) &\AL
\end{array} \bigg]
& = \begin{array}{rlrl} 
(((v_3) &\VL& v_{13}) p_2) &\VL \\
(((v_3) &\AL& v_{23}) p_1) &\AL
\end{array} \notag \\ 
&- ((v_3)o_{13,2|23,1}) - (((v_3)v_{13}) p_2) - (((v_3)v_{23}) p_1) ~, 
\\
\mathcal{P} \left[ \begin{array}{lrrrr}
( &(v_1) \VL&(v_2) \VL&         & p_3) \VL \\
((&(v_1) \AL&      \IL& v_{13}) & p_2) \TL \\
((&         &(v_2) \AL& v_{23}) & p_1) \AL
\end{array} \right]
&= 
\begin{array}{lrrrr}
( &(v_1) \VL&(v_2) \VL&         & p_3) \VL \\
((&(v_1) \AL&      \IL& v_{13}) & p_2) \TL \\
((&         &(v_2) \AL& v_{23}) & p_1) \AL
\end{array}
- ((v_1)o_{2,3|13,2}) - ((v_2)o_{2,3|23,1}) \notag \\ 
& - ((v_1)(v_2)p_3) - (((v_1)v_{13})p_2) - (((v_2)v_{23})p_1) ~.
\end{align*}
The meaning of the index structure of $o_{i,j|k,l}$ is obvious from
the first equation; we have to save this type of a primitive element
for later use in the second and third equation. The third equation,
for example, can be rewritten as the following decomposition of
\eqref{VVP} into a sum of rooted trees:
\[
\epsfig{file=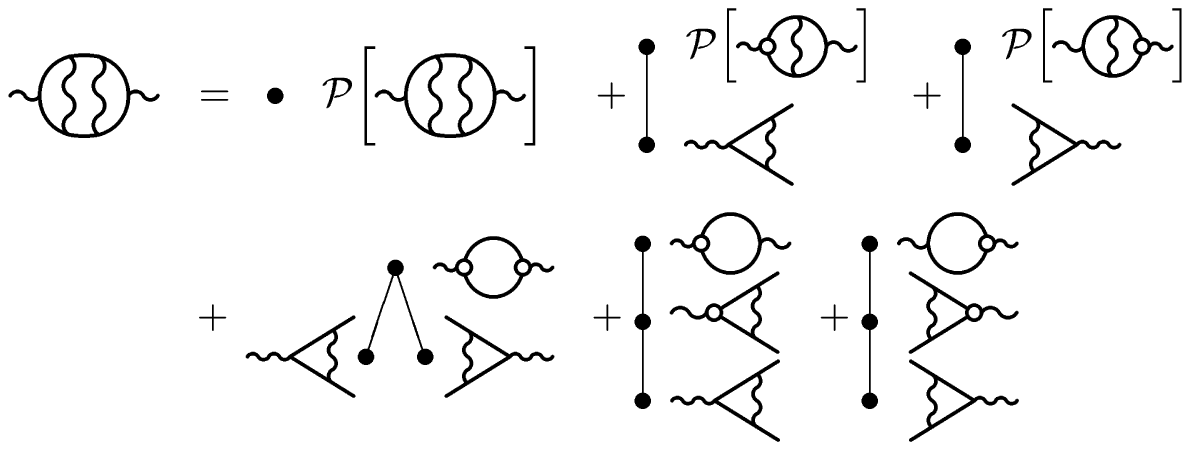}
\]

\section{Renormalization}
\label{R}

We have already mentioned that the antipode recovers the combinatorics
of the forest formula. In this section we will make this statement
explicit. In fact the antipode reproduces precisely the forest formula
if we allow for a deformation of our Hopf algebra by a renormalization
map $R$. That map $R$ should be considered as the projection onto the
divergent part of the integral encoded in the Feynman graph. In terms
of a regularization parameter $\epsilon$, those integrals deliver a
Laurent series, and $R$ is any projection of the Laurent series which
preserves the divergent part.  Renormalization schemes differ in the
way they handle the finite part in $\epsilon \to 0$. A special
scheme is BPHZ renormalization where the projection is given by Taylor
expansion of the integrals with respect to the external momenta.

We now enlarge our algebra $\mathcal{A}$ by copies $R[X]$ for each 
element $X \in \mathcal{A}$, subject to the convention $R[e]=e$. The
Hopf algebra definitions are now modified as follows:
\allowdisplaybreaks[4]
\begin{align}
\Delta[X] &:= e \otimes X + \tsum_U \Big\{ \tprod_{i \in U} \! R[X_i] 
\otimes X/ \!\tprod_{i \in U} \! X_i \Big\}~, \tag{$\ref{delta_P}_R$}
\\*
\Delta[R[X]] &= \Delta[X] \qquad \mbox{or} \tag{$\ref{delta_P}_R'$}
\label{DR1}
\\*
\Delta[R[X]] &= (\id \otimes R) \circ \Delta[X] ~, 
\tag{$\ref{delta_P}_R''$} \label{DR2}
\\
S[X] & = -X - m \circ (\id \otimes S) \circ P_2 \Delta [X]~,
\qquad \forall ~\mathrm{iPW}~X \in \mathcal{A}~, \tag{$\ref{anti}_R$}
\\
S[R[X]] & = -R[X - m \circ (S \otimes \id) \circ P_2 \Delta [X]]~,
\qquad \forall ~\mathrm{iPW}~X \in \mathcal{A}~, \tag{$\ref{anti}_R'$}
\end{align}
It turns out that a general $R$ spoils several Hopf algebra axioms,
somewhat depending on whether we prefer \eqref{DR1} or \eqref{DR2}. If
we choose \eqref{DR1} then counit, coassociativity and right antipode
axiom are given up, for the choice \eqref{DR2} we keep coassociativity
but violate the right counit and right antipode axioms. 

Let us explore the  left antipode axiom:
\begin{align*}
0 \sim m \circ (S \otimes \id) \circ \Delta[X]
&= m \circ (S \otimes \id) \circ (e \otimes X + R[X] \otimes e + P_2
\Delta[X]) 
\\
&= (\id -R)[X+ m \circ (S \otimes \id) \circ  P_2 \Delta[X]]
\\
=: (\id -R) [\bar{X}]
&= (\id-R) \Big[X + \tsum_U \Big\{ \tprod_{i \in U}  \big\{{-}R[\bar{X}_i]
\big\} \;\big\{ X/\tprod_{i \in U} X_i \big\} \Big\} \Big]~,
\\
R[\bar{X}_i] := -S[R[X_i]] & \equiv R[X_i+ m \circ (S \otimes \id)
\circ  P_2 \Delta[X_i]] ~.
\end{align*}
We see that we obtain a recursion formula for the determination of
$\bar{X}$, and this is precisely Bogolyubov's recursion formula
\cite{bs} of renormalization! That recursion formula has an explicit
solution, the forest formula of Zimmermann \cite{z}. In view of the
general results of \cite{k3} it is not surprising that Feynman graphs
can be regarded as elements of a Hopf algebra. It is however a deep
message that the antipode of this ($R$-deformed) Hopf algebra
implements renormalization. And the story continues with the discovery
by Connes and Kreimer of a connection between the renormalization Hopf
algebra and the diffeomorphism group of spacetime \cite{ck}. We can 
expect this subject to become one of the most promising activities in 
theoretical physics.  
\\[2ex] 
\emph{Acknowledgements:} I would like to thank Daniel Kastler for the
invitation to write this contribution as well as for the invitation to
come to Marseille. I am grateful to Dirk Kreimer for numerous
discussions and advice concerning overlapping divergences. 
Finally, I would like to thank Thomas Krajewski for the ongoing 
collaboration on this and other topics in noncommutative geometry.

\end{document}